\documentclass[10pt,conference,letterpaper]{IEEEtran}
\IEEEoverridecommandlockouts
\usepackage{cite}
\usepackage{amsmath,amssymb,amsthm}
\usepackage{bbm}
\usepackage{algpseudocode}
\usepackage{graphicx}
\usepackage{textcomp}
\usepackage{mathtools}
\usepackage{xcolor}
\usepackage{algorithm}
\usepackage{mathcommand}
\usepackage{bm}

\def\BibTeX{{\rm B\kern-.05em{\sc i\kern-.025em b}\kern-.08em
    T\kern-.1667em\lower.7ex\hbox{E}\kern-.125emX}}
    
\usepackage{url}
\usepackage{booktabs,threeparttable}
\usepackage{multirow,multicol}

\usepackage[caption=false]{subfig}
\usepackage{tikz}
\usetikzlibrary{shapes.arrows}
\usepackage{pgfplots}
\pgfplotsset{
compat=1.3,
legend style={font=\footnotesize, fill opacity=0.7,  draw opacity=1, text opacity=1, draw=white!15!black, legend cell align=left, align=left}, 
width=6cm, 
height=6cm,
yminorticks=false,
xminorticks=false,
title style={font=\small},
tick style={color=black},
tick label style={font=\small},
}
\usepgfplotslibrary{colorbrewer}
\usepgfplotslibrary{groupplots}
\usepgfplotslibrary{fillbetween}
\usetikzlibrary{decorations.pathreplacing}
\usetikzlibrary{plotmarks}
\usetikzlibrary{spy}
\usepackage[]{fancyhdr}
\usetikzlibrary{patterns}
\pgfplotsset{every tick label/.append style={font=\footnotesize}}

\usepackage{tikz}
\usetikzlibrary{shapes, arrows, positioning}

\usepackage{glossaries}
\newacronym{aoi}{AoI}{age of information}
\newacronym{api}{API}{alternate policy iteration}
\newacronym{ack}{ACK}{acknowledgment}

\newacronym{cmab}{CMAB}{contextual multi-armed bandit}
\newacronym{cps}{CPS}{cyber-physical system}

\newacronym{dt}{DT}{digital twin}

\newacronym{goc}{GoC}{goal-oriented communication}

\newacronym{jpo}{JPO}{joint policy optimization}

\newacronym{marl}{MARL}{multi-agent reinforcement learning}
\newacronym{mdp}{MDP}{Markov decision process}

\newacronym{ne}{NE}{Nash equilibrium}

\newacronym{pomdp}{POMDP}{partially observable Markov decision process}
\newacronym{poa}{PoA}{price of anarchy}

\newacronym{rl}{RL}{reinforcement learning}

\newacronym{voi}{VoI}{value of information}

\newmathcommand{\Esub}[2]{\mathbb{E}_{#1}\left[ #2 \right]} 
\newcommand{\mc}[1]{\mathcal{#1}}   
\DeclareMathOperator*{\argmax}{arg\,max}    

\algtext*{EndWhile}
\algtext*{EndIf}
\algtext*{EndFor}

\def \fwidth{0.96\columnwidth}
\def \fheight {0.45\columnwidth}

\definecolor{color0}{HTML}{00429D}
\definecolor{color1}{HTML}{844D99}
\definecolor{color2}{HTML}{C3608E}
\definecolor{color3}{HTML}{EF8078}
\definecolor{color4}{HTML}{FFB047}
\definecolor{darkslategray38}{RGB}{38,38,38}
\definecolor{darkslategray88}{RGB}{88,88,88}
\definecolor{lightgray158}{RGB}{158,158,158}

\newtheorem{theorem}{Theorem}

\newtheorem{definition}{Definition}
\newtheorem{lemma}{Lemma}[theorem]

\newtheorem{corollary}{Corollary}[theorem]

\makeatletter

\title{On the Optimality of Implicit Feedback in Markov Decision Processes over Imperfect Links}
\author{\IEEEauthorblockN{Romain Planchon\IEEEauthorrefmark{1}, Federico Chiariotti\IEEEauthorrefmark{2}, Federico Mason\IEEEauthorrefmark{2}, Andrea Zanella\IEEEauthorrefmark{2}, and Deniz G\"u{}nd\"u{}z\IEEEauthorrefmark{3}}
\IEEEauthorblockA{\IEEEauthorrefmark{1}T\'el\'ecom Paris, France (email: romain.planchon@telecom-paris.fr) \\}
\IEEEauthorblockA{\IEEEauthorrefmark{2}Department of Information Engineering, University of Padova, Italy (emails:\{name.lastname\}@unipd.it) \\}
\IEEEauthorblockA{\IEEEauthorrefmark{3}Department of  of Electrical and Electronic Engineering, Imperial College London, UK (email: d.gunduz@imperial.ac.uk) \\}\vspace{-0.5cm}
\thanks{We acknowledge funding from the UKRI for the projects AI-R (ERC Consolidator Grant, EP/X030806/1) and Smart Networks and Services Joint Undertaking (SNS JU) projects 6G-GOALS (Grant Agreement No. 101139232) and MAGIC-6G (Grant Agreement No. 101292933) under the EU’s Horizon Europe research and innovation program.}}

\begin{document}

\maketitle

\begin{abstract}
Goal-oriented communication protocols are explicitly designed to support the objectives of target applications, such as the monitoring and control of remote processes, while minimizing data traffic. This study considers a transmitter node that observes the environment state and sends it to a remote agent, which then performs control actions.
This problem can be modeled as a remote \gls{pomdp}, a subclass of decentralized \glspl{pomdp}. 
Existing studies limit transmissions through communication costs, but neglect the effect of channel impairments and non-ideal or delayed feedback.
In this work, we study the optimal joint transmission and control policies over lossy forward channels without explicit acknowledgments. We introduce \emph{revealing} policies, under which the receiver's action, observed directly by the transmitter, discloses the outcome of the forward transmission, keeping the two agents synchronized without a feedback channel.
We derive sufficient conditions for the optimality of time-invariant policies in this implicit feedback regime for general \glspl{pomdp}, along with strictly weaker sufficient conditions in the special case of contextual bandits.
Our simulations over a family of remote \glspl{pomdp} show that these conditions are conservative, and revealing policies are optimal over a substantially wider parameter range than the conditions warrant.
\end{abstract}


\glsresetall

\section{Introduction}

\begin{tikzpicture}[remember picture, overlay]
      \node[draw,minimum width=3in] at ([yshift=-1cm]current page.north)  {This manuscript is curently under review.};
\end{tikzpicture}
The sixth generation of cellular networks (6G) foresees an ever tighter integration between the physical and digital worlds~\cite{pennanen20256g}.
Enabling the control of cyber-physical systems via emerging paradigms such as digital twins~\cite{khan2022digital} is a major expected benefit of 6G for industrial automation and robotics.
However, the level of safety and efficiency that these applications demand requires throughput, latency, and reliability that already approach the network's Shannon capacity limits~\cite{zhang2026beyond}.

The \gls{goc} paradigm allows system designers to sidestep the issue by jointly optimizing the end applications and the network protocols: delivering information about the state of the system only when it is needed by the control or estimation algorithm substantially reduces the burden on the network without affecting the performance of the end applications~\cite{shi2023task, gunduz2023beyond, gunduz2023timely}.
On the other hand, \gls{goc} optimization is often difficult, as the usefulness of a single update may depend on the entire history of transmitted messages and the current state of the environment~\cite{luo2026role, bismut1972example}.

To optimize two-agent \gls{goc} systems, we can take advantage of the remote \gls{pomdp} model~\cite{talli2025pragmatic,li2026from}, capturing scenarios in which a \emph{transmitter} acquires the state of an environment and decides whether to send an update to a \emph{receiver}, which cannot observe the environment directly, but can perform control actions on it. Explicit communication in a remote \gls{pomdp} happens over a forward channel, over which the transmitter sends updates to the receiver, and a backward channel, over which the receiver sends \gls{ack} packets.
A key feature of remote \glspl{pomdp} is that the absence of an update provides \emph{implicit information} to the receiver, which can infer that the transmitter chose to remain silent, and refine its own state estimate through Bayesian reasoning~\cite{talli2025pragmatic, bozkurt2025generalized}. 

If either of the communication channels is ideal, the transmitter always knows exactly what observations have been delivered to the receiver and, consequently, what the receiver knows. In these conditions, the globally optimal policy in a remote \gls{pomdp} is always \emph{time-invariant}~\cite{talli2025pragmatic}: this greatly reduces the complexity of finding it and allows it to be stored in finite memory. This is not true for the wider class of decentralized \glspl{pomdp}, of which remote \glspl{pomdp} represent a special case and which can model a wide class of distributed control problems~\cite{goldman2004decentralized}.
In the general case, the optimal strategy always exists~\cite{nayyar2013decentralized}, but it may not be time-invariant, i.e., it may depend on the full history of past observations~\cite{amato2013decentralized}.

An ideal backward channel is necessary to maintain an \emph{information hierarchy} between the transmitter and receiver, i.e., for the former to always know exactly what the latter knows. This hierarchy can be exploited to simplify the optimization problem~\cite{nayyar2019common}, but real communication scenarios must consider feedback packet losses or delays due to, e.g., wireless access and propagation issues. The propagation delay of the backward channel is a major issue in real-time control systems~\cite{cao2023age}, and particularly in industrial automation networks with strict latency requirements~\cite{yu2022industrial}, and delays in the reception of \glspl{ack} can affect the sampling strategy in monitoring tasks~\cite{liyanaarachchi2024role}. In extreme scenarios, such as underwater and satellite networks, the backward channel may be very inconvenient due to high delay and frequent packet losses~\cite{campagnaro2014on}, or even entirely absent, such as in covert communication scenarios~\cite{lucia2021covert}. To the best of our knowledge, realistic remote \glspl{pomdp} with imperfect forward and backward channels remain intractable, and the existence of optimal time-invariant policies is open. This work is a first step toward solving this fundamental problem.

We consider a remote \gls{pomdp} with imperfect forward and backward channels and assume that the transmitter can observe the receiver's actions at each time step.
While the considered scenario can be optimized through the general decentralized \glspl{pomdp} paradigm~\cite{tavafoghi2022unified}, the resulting policies will be history-dependent. 
Hence, we introduce the concept of \emph{revealing policies}, i.e., communication and control policies that maintain the information hierarchy between transmitter and receiver without \glspl{ack} by using the receiver's actions as implicit feedback.
We prove that revealing policies have time-invariant equivalents that can be computed using frameworks for standard remote \glspl{pomdp} and stored in finite memory. We analyze their performance against their counterparts with an explicit \gls{ack} mechanism over an ideal backward channel. 

Our main result shows that, under a wide range of system parameters, the optimal transmission and control policies are always revealing, making the backward channel unnecessary and allowing for optimization considering only time-invariant policies.
We also extend our results in the case of remote \glspl{cmab}, in which the actions performed by the receiver do not affect the environment's state.
For this class of problems, which includes remote estimation as a special case~\cite{pase2022rate,gupta2023remote}, the set of system parameters under which revealing policies are optimal is even wider.

The contributions of this work are summarized as follows:
\begin{itemize}
    \item We provide a rigorous analysis of the remote \gls{pomdp} problem with imperfect communication, extending existing results with ideal channels to more general cases;
    \item We introduce \textit{revealing} communication and control policies, which can exploit implicit feedback to operate over an unreliable channel without \glspl{ack}, proving that they have time-invariant equivalents;
    \item We derive a sufficient condition for the global optimality of time-invariant revealing policies, proving a stronger result in the special case of remote \glspl{cmab};
    \item We derive upper and lower performance bounds for generic optimal policies, showing that the gap between the optimal policy with perfect \glspl{ack} and an intuitive heuristic is small over a family of remote \glspl{pomdp}.
\end{itemize}
The rest of this paper is organized as follows: first, Sec.~\ref{sec:related} presents the state of the art on remote \glspl{pomdp}; Sec.~\ref{sec:system} then introduces the remote \glspl{pomdp} model.
Sec.~\ref{sec:theory} presents our theoretical results, which are verified by simulation in Sec.~\ref{sec:simulation}. Finally, Sec.~\ref{sec:conc} concludes the paper.


\section{Related Work}\label{sec:related}

The decentralized \gls{pomdp} problem~\cite{goldman2004decentralized} has been widely studied in the communication and control literature due to its highly general nature, which includes a number of useful subclasses.
In a decentralized \gls{pomdp}, multiple agents can both act in a shared environment and communicate part of their observations to other agents~\cite{nayyar2013decentralized}, yielding complex information structures~\cite{malikopoulos2023on}.
In the most general case, while some elements of the knowledge of the agents may be reduced to a belief distribution, the full history of observations and communications must be maintained in each agent's memory to avoid losing useful information~\cite{nayyar2019common}.
Although many Bayesian reasoning strategies have been developed to solve decentralized \gls{pomdp} problems~\cite{bozkurt2025generalized}, the lack of any recognized information structure significantly hinders their analysis and optimization in practical use cases~\cite{sudhakara2024optimal}.

The remote \gls{pomdp} problem is a subclass of decentralized \glspl{pomdp} that considers only two agents with distinct roles, often admitting a strong information hierarchy. The transmitter is constrained to only communicate and never act, while the receiver can act, but communication is limited to acknowledging the successful reception of an update. These assumptions reduce the degrees of freedom of the problem.
In particular, under ideal communication, the strategy space coincides with the set of time-invariant communication and control policies, whose input state can be stored in finite memory. 
Although finding the optimal joint policies for this scenario is still computationally hard~\cite{talli2025pragmatic}, efficient heuristics that can achieve near-optimal performance~\cite{salimnejad2024real} or local optimality guarantees~\cite{chiariotti2025solving} have been developed.
Particular subclasses of remote \glspl{pomdp}, such as linear systems~\cite{lipsa2011remote}, admit even more efficient solutions, which have been thoroughly investigated in the recent literature, considering both continuous~\cite{molin2017event} and discrete~\cite{salimnejad2024real} state spaces. In recent years, remote \glspl{pomdp} with capacity-limited channels have also been analyzed, and several learning-based solutions have been developed to encode information or schedule transmissions~\cite{wang2020learning,tung2021effective}.

The remote estimation problem, in which the receiver does not affect the evolution of the process, represents a subclass of remote \glspl{pomdp} that has attracted significant interest.
In particular, in this scenario, maximizing performance in the worst case rather than on average yields a much simpler optimization, as the resulting problem has an open-loop optimal strategy~\cite{gagrani2021worst}.
Capacity-limited remote estimation problems have been largely studied considering both continuous~\cite{dey2014remote,khina2019tracking} and discrete~\cite{vilni2026realpull,vilni2026real} state spaces.
However, the complexity of these problems has been shown to be the same as for the general case of remote \glspl{pomdp}~\cite{talli2025pragmatic}.

The optimization of remote \glspl{pomdp} with unreliable feedback, which is the focus of the present work, has seen limited development in the literature.
To the best of our knowledge, aside from some early work considering Kalman filter estimation of a linear system~\cite{nourian2014optimal}, only one recent work~\cite{soleymani2026remote} considered the remote \gls{pomdp} case with imperfect \glspl{ack}.
Even this work focused on a simple estimation scenario, optimizing packet scheduling for a linear dynamic system with Gaussian noise that affects the state evolution.
Therefore, the case of arbitrary remote \glspl{pomdp} with an imperfect or missing backward channel remains unsolved in the current literature.

\section{System Model}\label{sec:system}

In the following, we denote sets using calligraphic letters, e.g., $\mc{S}$, while vector $(x_0,\ldots,x_t)$ is shortened as $x_{0:t}$. Finally, we write $\Phi(\mc{S})$ to indicate the probability simplex over $\mc{S}$. 

\subsection{The Remote POMDP Model}
We consider an infinite-horizon remote \gls{pomdp}, in which a receiver agent must control a process without being able to observe the state directly~\cite{talli2025pragmatic}, while a transmitter agent can see the process state and the receiver's actions, and is allowed to communicate the state of the process to the receiver over a packet erasure channel.
Each transmission incurs a communication cost and, if not erased, elicits an \gls{ack} packet, which is cost-free but also subject to potential erasures.
\begin{definition}
    A remote \gls{pomdp} model is characterized by the tuple $\langle \mc{S},\mc{A},P,r,\gamma,\beta,\varepsilon,\varphi\rangle$, whose elements are:
\begin{itemize}
  \item The state space $\mc{S}$, with cardinality $S$;
  \item The receiver's action space $\mc{A}$, with cardinality $A$;
  \item The transition function $P:\mc{S}\times\mc{A}\to\Phi(\mc{S})$, which provides the distribution of the next state $s^\prime\in \mc{S}$ given that the receiver takes action $a\in\mc{A}$ in state $s\in \mc{S}$ ;
  \item The task reward function $r:\mc{S}\times\mc{A}\times\mc{S}\to[0,1]$, which denotes the reward for transition $\langle s,a,s'\rangle$;
  \item The exponential discount factor $\gamma$ over future rewards;
  \item The communication cost $\beta$ of packet transmissions;
  \item The forward channel erasure probability $\varepsilon$;
  \item The \gls{ack} channel erasure probability $\varphi$.
\end{itemize}
\end{definition} 

Time is discretized into steps and, at each step $t$, the transmitter observes the receiver's previous action $a_{t-1}$ and the current state $s_t$ of the process, then decides a communication action $c_t\in\{0,1\}$, which may be to transmit the state value to the receiver ($c_t=1$) or remain silent ($c_t=0$).

We assume that the receiver cannot distinguish between erasures or skipped transmissions. We then denote by $m_t\in\mc{M}$ the decoding outcome at the receiver at time slot $t$. In case of a successful transmission, $m_t=s_t$. Otherwise, $m_t$ is set to a null symbol $\chi$, so that $\mc{M}=\mc{S}\cup\{\chi\}$. For the sake of simplicity, in the following we refer to $m_t$ as the \textit{received message} even when no update is actually received. If $m_t\neq \chi$, the receiver acknowledges the correct reception of the state update by sending an \gls{ack} to the transmitter through the backward channel. Otherwise, it does not return any message. The receiver then takes an action $a_t\in\mc{A}$, which affects the evolution of the process according to the transition probability function $P$, yielding the immediate reward at time $t$
\begin{equation} 
R(s_t,c_t,a_t,s_{t+1})=r(s_t,a_t,s_{t+1})-\beta c_t.
\end{equation}

We denote the explicit feedback observation at the transmitter by $f_t\in\{0,1\}$, with $f_t=1$ if the transmitter receives an \gls{ack}, and $f_t=0$ otherwise, that is, if either the update or the feedback packet was erased. Note that these two events are indistinguishable to the transmitter when considering only the information from the data channels. However, as explained in the following, the transmitter can also infer the outcome of the transmission by observing the action $a_t$ taken by the receiver. 

\subsection{Transmitter and Receiver Histories and Policies}

The possibility of erasures in the forward and \gls{ack} channels completely changes the nature of  the problem compared to that addressed in~\cite{salimnejad2024real,talli2025pragmatic}, where the communication channels were ideal ($\varepsilon=\varphi=0$). Specifically, erasures introduce uncertainty in the transmitter's knowledge of the information possessed by the receiver. This prevents the transmitter from maintaining a compact representation of the receiver's belief state, ultimately disrupting the information hierarchy between the transmitter and receiver that underpins prior solutions.

We assume that, at the beginning of step $t$, the transmitter knows the current process state $s_t$ and the result of the previous interaction, which includes the transmission decision $c_{t-1}$, the action $a_{t-1}$ performed by the receiver, and the feedback observation $f_{t-1}$. We collect these observed variables in the tuple $o_{\text{tx},t}=\langle c_{t-1},a_{t-1},f_{t-1},s_t\rangle$. The full knowledge of the transmitter at time $t$ is then given by the cumulative history of these variables, which we indicate as $H_{\text{tx},0:t}=\langle c_{0:t-1},a_{0:t-1},f_{0:t-1},s_{0:t}\rangle$, taking values in the set $\mc{H}_{\text{tx},t}=\{0,1\}^t\times\mc{A}^t\times\mc{F}^t\times\mc{S}^{t+1}$.
The transmitter can then implement a time-dependent communication policy, denoted as $\pi_{\text{tx},t}:\mc{H}_{\text{tx},t}\to\Phi(\{0,1\})$, that maps each possible history to a probability distribution over the action space.

Similarly, we denote the new information available at the receiver at the beginning of time step $t$ as $o_{\text{rx},t}=\langle m_{t-1},a_{t-1}\rangle$.
The knowledge of the receiver is represented as the history tuple $H_{\text{rx},0:t}=\langle m_{0:t-1},a_{0:t-1}\rangle$, which belongs to space $\mc{H}_{\text{rx},t}=\mc{M}^t\times\mc{A}^t$.
The receiver knows both $H_{\text{rx},0:t}$ and the received message $m_t$, so its time-dependent control policy is  $\pi_{\text{rx},t}:\mc{H}_{\text{rx},t}\times\mc{M}\to\Phi(\mc{A})$.

\begin{definition}\label{def:history_dep}
    The joint policy space $\Pi$ is the combination of the two policy sets containing all possible transmitter and receiver policies, respectively:
    \begin{equation}
      \Pi=\left(\mc{H}_{\text{tx},t}\to\Phi(\{0,1\})\right)\times\left(\mc{H}_{\text{rx},t}\times\mc{M}\to\Phi(\mc{A})\right).  
    \end{equation}
\end{definition}

\subsection{Value Functions}
Our goal is to find the combination of communication and control policies $\bm{\pi}=\langle\pi_{\text{tx}},\pi_{\text{rx}}\rangle \in \Pi$ that maximizes the long-term performance, quantified by the value function 
\begin{equation}\label{eq:value_tx}
\begin{aligned}
  V_{s}^{\pi_{\text{tx}},\pi_{\text{rx}}}(h_{\text{tx}},h_{\text{rx}})=\mathbb{E}\Bigg[\sum_{\tau=t}^{\infty}\gamma^{\tau-t}\sum_{\mathclap{a\in\mc{A},c\in\{0,1\}}}R\big(s_{\tau},\pi_{\text{tx},\tau}(c|H_{\text{tx},0:\tau}),\\
  \pi_{\text{rx},\tau}(a|H_{\text{rx},0:\tau},m_t),s_{\tau+1}\big)\big|s_t\!=\!s,H_{\text{tx},0:t}\!=\!h_{\text{tx}},H_{\text{rx},0:t}\!=\!h_{\text{rx}}\Bigg].
\end{aligned}
\end{equation}

In the rest of the paper, for the sake of readability, we will denote the expected long-term value of a joint policy $\bm{\pi}$ starting at time $0$ and with a state distribution $\mu_0$ as $V^{\bm{\pi}}= \Esub{s\sim\mu_0}{V_{s,0}^{\bm{\pi}}(s,\varnothing)}$.

\section{Theoretical Analysis}
\label{sec:theory}

We first define a condition that establishes whether the two agents maintain an information hierarchy.
If the transmitter knows exactly what the receiver knows, the two agents can exploit the Markov property to find an equivalent time-invariant representation of their knowledge over the environment's state. 
In particular, under these conditions, the full knowledge of the agents depends only on the current state (which is known only to the transmitter), the last successfully transmitted state, and the time since the last successful reception of the update, i.e., \gls{aoi} $\Delta_t$~\cite{yates2021age}, defined as
\begin{equation}
    \Delta_t=t-\sup\left\{\tau\in\{0,\ldots,t-1\}:m_{\tau}=s_{\tau}\right\}.
\end{equation}
The transmitter's history at time $t$ is then represented by the tuple $\langle s_t,\Delta_t,s_{t-\Delta_t}\rangle$ and the receiver's history at time $t$ by $\langle\Delta_t,s_{t-\Delta_t}\rangle$.
Note that this representation of the knowledge of the two agents is \emph{time-invariant}, as the number and nature of the elements in the tuples do not vary with $t$. While this is not necessary to find the global optimum, which can be computed by applying the \gls{jpo} algorithm~\cite[Alg. 3]{talli2025pragmatic}, reducing the search space significantly improves its computational efficiency, and optimal time-invariant policies can be stored in finite memory without approximations.

\subsection{Time-Invariant Policies}

If the two agents maintain the information hierarchy, they can exploit time-invariant policies, which present a finite input space and thus can be stored in finite memory. 
\begin{definition}\label{def:time_inv}
    The time-invariant policy space $\Pi^{\text{TI}}\subset\Pi$ is the combination of the two policy sets containing all possible transmitter and receiver policies that reset the nodes' memories after each successful transmission:
    \begin{equation}
      \Pi^{\text{TI}}=\left(\mc{S}\times\mc{S}\times\mathbb{N}\to\Phi(\{0,1\})\right)\times\left(\mc{M}\times\mc{S}\times\mathbb{N}\to\Phi(\mc{A})\right).  
    \end{equation}
\end{definition}
Restricting the problem to time-invariant policies raises optimality concerns, as the optimal solution may not be time-invariant.
In the following, we will define sufficient conditions for time-invariant policies to have optimality guarantees, including the need for the agents to be \emph{synchronized}.

\begin{definition}
  Given the joint policy $\bm{\pi}_t=\langle\pi_{\text{tx}, t},\pi_{\text{rx}, t}\rangle$ at time $t$, a transmitter-side history $h_{\text{tx}}$ is \emph{synchronized} if there exists one receiver-side history $h_{\text{rx}}$ such that
  \begin{equation}
    \text{Pr}\left[H_{\text{rx},0:t}=h_{\text{rx}}|H_{\text{tx},0:t}=h_{\text{tx}},\pi_{\text{tx}, t},\pi_{\text{rx}, t}\right]=1.
  \end{equation}
\end{definition}

In the general case, synchronization depends on the joint policy of the transmitter and receiver, as they make take actions that break synchronization, causing the information hierarchy between the transmitter and receiver to break down.

\begin{theorem}\label{thm:time_invariant_equiv}
If all possible transmitter-side histories are synchronized under the optimal policy $\bm{\pi}^*=\argmax_{\bm{\pi}\in\Pi}V^{\bm{\pi}}$, time-invariant policies are globally optimal:
\begin{equation}
     \max_{\bm{\pi} \in \Pi} V^{\bm{\pi}}  = \max_{\bm{\pi}' \in \Pi^{\text{TI}}} V^{\bm{\pi}'}.
\end{equation}
\end{theorem}
\begin{IEEEproof}
Consider a case in which the optimal policy only uses the time-invariant information starting from time $T>0$. For an arbitrary $h_{\text{rx}}\in\mc{H}_{\text{rx},t}$, we consider the steps between $T-\Delta_t$ and $T$. Thanks to the policy-independence property of beliefs~\cite{tavafoghi2022unified}, we can find a joint policy that only uses the time-invariant tuple between $T-\Delta_t$ and $T$ and maintains the same long-term value, as future rewards are not affected by information before time $T$. We can thus proceed by backward induction until we reach step $0$, and the joint policy has been substituted by a policy in $\Pi^{\text{TI}}$ with no change in the reward. We now consider the initial assumption. The difference between the optimal policy and a policy that is identical up to time $T$, but changes from step $T$ onward, is bounded by $\gamma^T(1-\gamma)^{-1}$, and the time-invariant policy is equivalent to the latter. By increasing the value of $T$, we can reduce the optimality gap between the two policies at will, and taking the limit for $T\to\infty$, the optimality gap reaches $0$.
\end{IEEEproof}

Theorem~\ref{thm:time_invariant_equiv} allows us to restrict our optimization problem to the time-invariant policy space, as long as we can guarantee synchronization under the optimal policy. 
If either the communication or the backward channel is ideal, i.e., if $\varepsilon=0\vee\varphi=0$, this is always the case, as the outcome of each transmission is known, ensuring synchronization for any joint policy.

\subsection{Nash Equilibria}
\glsreset{ne}

As the complexity of \gls{jpo} is still a significant issue, we can use sub-optimal heuristics as an alternative.
For instance, \gls{api}~\cite[Alg. 2]{talli2025pragmatic} reaches local optimality in remote \glspl{pomdp} by iterative optimization, adopting a game theoretical approach.
The two agents of our model share the same reward and can be thought of as players in a \emph{fully cooperative game}, a subclass of \emph{exact potential games}.
Therefore, local optimality corresponds to the concept of \gls{ne} in this class of games, and we can consider the properties of \glspl{ne} in $\Pi^{\text{TI}}$.

\begin{definition}
A joint policy $\bm{\pi}=\langle\pi_{\text{tx}},\pi_{\text{rx}}\rangle$ is a \gls{ne} over policy space $\Pi$ if no unilateral deviation from either agent improves the policy's expected value, that is
\begin{equation}
  \nexists\, \pi_{\text{tx}}'\in\Pi_{\text{tx}}:V^{\pi_{\text{tx}}',\pi_{\text{rx}}}>V^{\bm{\pi}}\wedge  \nexists\, \pi_{\text{rx}}'\in\Pi_{\text{rx}}:V^{\pi_{\text{tx}},\pi_{\text{rx}}'}>V^{\bm{\pi}}.
\end{equation}
\end{definition}
We denote the set of \glspl{ne} in $\Pi$ as $\mc{E}(\Pi)$. As the problem is an exact potential game, the \gls{poa}~\cite{koutsoupias2009worst} $\sigma(\Pi)$ is the difference between the global optimum, which is always an \gls{ne}, and the worst possible \gls{ne}, and we have
\begin{equation}\label{eq:poa}
    \sigma(\Pi)=\sup_{\bm{\pi}\in\mc{E}(\Pi)}V^{\bm{\pi}}-\inf_{\bm{\pi}'\in\mc{E}(\Pi)}V^{\bm{\pi}'}.
\end{equation}

\begin{theorem}\label{thm:ne_time_inv}
If the time-invariant joint policy $\bm{\pi}\in\Pi^{\text{TI}}$ is an \gls{ne} over $\Pi^{\text{TI}}$, and all possible transmitter-side histories are synchronized under it, then $\bm{\pi}$ is an \gls{ne} over $\Pi$, i.e., in the full-history dependent game.
\end{theorem}
\begin{IEEEproof}
In order for the joint policy to be an \gls{ne}, neither agent must have regrets, i.e., it must not be able to improve the reward by unilaterally changing its actions.
Since all possible transmitter-side histories are synchronized, the Markov property allows us to say that any element of the history aside from the time-invariant state is irrelevant, i.e., it does not affect future rewards. Therefore, any history-dependent unilateral change in the policy will not improve the overall reward, and the \gls{ne} over $\Pi^{\text{TI}}$ is also an \gls{ne} over $\Pi$.
\end{IEEEproof}

\begin{corollary}\label{cor:poa}
    If all possible transmitter-side histories are synchronized, $\sigma(\Pi^{\text{TI}})\leq\sigma(\Pi)$.
\end{corollary}
\begin{IEEEproof}
By Theorem~\ref{thm:ne_time_inv}, all \glspl{ne} in $\Pi^{\text{TI}}$ are \glspl{ne} in $\Pi$, i.e., $\mc{E}(\Pi^{\text{TI}})\subseteq\mc{E}(\Pi)$. The bound $\sigma(\Pi^{\text{TI}})\leq\sigma(\Pi)$ then trivially follows from the definition of the \gls{poa} in~\eqref{eq:poa}.
\end{IEEEproof}

Corollary~\ref{cor:poa} ensures that running \gls{api} over $\Pi^{\text{TI}}$ preserves its local optimality properties even when considering the full policy space $\Pi$.
In particular, restricting the space of solutions to time-invariant policies may remove some low-reward equilibria, leading to a lower \gls{poa}.

We note that restricting to time-invariant policies does not, by itself, ensure that their input space is finite; the \gls{aoi} $\Delta_t$ may grow unboundedly under policies that transmit rarely. To store policies in finite memory and run \gls{api}, we therefore need to cap \gls{aoi} to a value $\Delta_{\max}$.
This truncation is asymptotically lossless; the performance gap introduced by the cap decays geometrically in $\Delta_{\max}$, and can be made arbitrarily small.

\subsection{Revealing Policies under Imperfect Communication}

In the rest of the section, we will restrict our analysis to deterministic policies without loss of optimality, and consider the general case in which neither the forward nor the backward channels are ideal.
As the optimality of time-invariant policies is guaranteed for synchronized policies, our aim is to ensure synchronization without the need for \glspl{ack}.
This can be done through implicit feedback: under some conditions, the receiver's actions may \emph{reveal} its knowledge to the transmitter, which will then be able to unequivocally identify the outcome of a transmission even without an \gls{ack}.

\begin{definition}
Under the restriction to deterministic policies, the receiver policy $\pi_{\text{rx}}$ is \emph{revealing} at history $h_{\text{rx}}$ and state $s$ if the action taken upon successful reception differs from that taken in case of erasure, i.e., $
\pi_{\text{rx}}(h_{\text{rx}}, s)\neq \pi_{\text{rx}}(h_{\text{rx}}, \chi)$.
\end{definition}

In cases in which the receiver policy is \emph{revealing}, receiving a message leads the receiver to select a different control action, and the transmitter can deduce $m_t$ from observing $a_t$.
On the other hand, if the transmitter does not send an update at step $t$, synchronization is maintained independently from $\pi_{\text{rx}}$, since the receiver will always get symbol $\chi$.

\begin{lemma}\label{lm:implicit_sync}
If transmissions up to time $t-1$ only occur in states and histories in which the receiver's policy is revealing, all possible transmitter-side histories up to time $t$ are synchronized for any value of $\varphi$.
\end{lemma}
\begin{IEEEproof}
We prove the Lemma by induction.
By the definition of the model, the initial transmitter history is synchronized, as the receiver history at $t=0$ is empty.
If history is synchronized up to time $\tau$, we have two possible cases for step $\tau+1$.
In the case in which $c_{\tau}=0$, the receiver must observe $m_{\tau}=\chi$, as there is no transmission, and the history is still synchronized.
On the other hand, if $c_{\tau}=1$, the receiver may observe either $s_{\tau}$ or $\chi$.
However, its action $a_{\tau}$ is different
in these two cases, as stated by the definition of revealing policies.
Since $a_{\tau}$ is part of the transmitter-side observation $o_{\text{tx},\tau}$, the transmitter can determine whether the message was received or not, maintaining history synchronization.
\end{IEEEproof}

We can then extend the definition to a space of \emph{revealing policies}, which maintain this property for any time step $t$:
\begin{definition}
\label{def:rev-policies}
The set of revealing policies $\Pi^{\text{RV}}\subset\Pi$ is:
\begin{equation}
\begin{aligned}
    \Pi^{\text{RV}}=\big\{&\langle\pi_{\text{tx}},\pi_{\text{rx}}\rangle\in\Pi:\pi_{\text{tx}}(h_{\text{tx}})=0\\
    &\vee\pi_{\text{rx}}(h_{\text{rx}}, s)\neq \pi_{\text{rx}}(h_{\text{rx}}, \chi),\ \forall s\in\mc{S},h_{\text{tx}},h_{\text{rx}}\big\}.
\end{aligned}
\end{equation}
\end{definition}

If the receiver policy is revealing whenever the transmitter updates the state to the receiver, we can eliminate the need for a backward channel while still ensuring synchronization. 
Hence, under any policy $\bm{\pi}\in\Pi^{\text{RV}}$, the system is indistinguishable from the case with $\varphi=0$, as the transmitter can always recover the value of $m_t$, and substituting $f_t$ with this value yields the same results of the ideal feedback case. 
\begin{lemma}
\label{lm:phi0-optimality}
     If the optimal policy with ideal feedback $\bm{\pi}_{\varphi=0}^{*}$ is revealing, it is also optimal for any $\varphi>0$.
\end{lemma}
\begin{IEEEproof}
    If we have $\bm{\pi}_{\varphi=0}^{*}\in\Pi^{\text{RV}}$, then the system while using the optimal policy is indistinguishable from the case with $\varphi=0$, which means that $\bm{\pi}_{\varphi=0}^{\star}$ is a valid policy for any $\varphi$.
    Moreover, $\forall \varphi,  V_\varphi^{\star} \leq V_{\varphi=0}^{\star}$, as a perfect feedback channel policy can always be played by discarding information.
\end{IEEEproof}

We note that Theorem~\ref{thm:time_invariant_equiv} ensures that we can compute the optimal policy with ideal feedback $\bm{\pi}_{\varphi=0}^{*}$ using the \gls{jpo} algorithm, while Lemma~\ref{lm:phi0-optimality} allows us to extend its optimality to any case $\varphi>0$ as long as $\bm{\pi}_{\varphi=0}^{\star}$ is revealing.
As the next step, we look for sufficient conditions under which \glspl{ne}, which correspond to local optima, are associated with revealing policies.
Under these conditions, policy $\bm{\pi}_{\varphi=0}^{\star}$, which is also a \gls{ne}, is optimal for any value of $\varphi$.

\begin{theorem}\label{thm:oneway_cond}
  If the joint policy $\bm{\pi}$ is an \gls{ne}, it must be revealing for any value of $\varphi$ if the remote \gls{pomdp} respects condition
  \begin{equation}\label{eq:beta_cond}
    \beta>\gamma\varepsilon(1-\varepsilon)(1-\gamma)^{-1},
  \end{equation}
\end{theorem}
\begin{IEEEproof}
The proof is presented as Appendix A.
\end{IEEEproof}

\begin{corollary}\label{cor:oneway_opt}
Under the condition in~\eqref{eq:beta_cond}, Theorem~\ref{thm:time_invariant_equiv} and Theorem~\ref{thm:ne_time_inv} hold in the general case.
\end{corollary}
\begin{IEEEproof}
 The corollary follows trivially from Lemma~\ref{lm:implicit_sync}: as revealing policies maintain synchronization, the conditions of the two theorems are verified.
\end{IEEEproof}

The result of Theorem~\ref{thm:oneway_cond} is sufficient for time-invariant policies to be optimal and for the \gls{poa} to be lower than in the full policy space.
This allows us to compute the optimal policies using any method developed for ideal \gls{ack} channels without any modifications, as observing the receiver's actions provides an ideal implicit feedback mechanism.

\begin{figure}[t]
\centering
\tikzstyle{every pin}=[
    fill=white,
    draw=black,
    font=\footnotesize,
] 

\begin{tikzpicture}
\begin{axis}[
    clip mode=individual,
    at={(0,0)},
    width=\fwidth,
    height=\fheight,
    xlabel={$\varepsilon$},
    ylabel={$\beta_{\min}(\varepsilon)$},
    xmin=0, xmax=0.5,
    ymin=0, ymax=4,
    ytick={0,1,2,3,4,5},
    legend pos=south east,
    ymajorgrids=true,
    xmajorgrids=true,
    grid style=dashed,
    legend style={overlay,legend columns=1, font=\scriptsize, anchor=south east, at={(1,1.05)}}
]

\coordinate (c0) at (axis cs:0,2);
\coordinate (c1) at (axis cs:0.1,0);
\coordinate (c2) at (axis cs:0.1,2);
\coordinate (c3) at (axis cs:0.08,2);
\coordinate (c4) at (axis cs:0.155,6);

\draw[semithick,darkslategray88] (c0) -- (c2);
\draw[semithick,darkslategray88] (c2) -- (c1);
\draw[darkslategray88,semithick] (c3) -- (c4);

\path[name path=bottom] (axis cs:0,0) -- (axis cs:0.1,0);

\path[name path=top] (c0) -- (c2);

\addplot[forget plot,color=gray!10] fill between [of=bottom and top, on layer=axis background];

\addplot[
    color=color0, semithick,
    mark=triangle*, mark repeat=40
    ]
    table[x=e,y=g9] {fig/fig_data/bounds.dat};
\addlegendentry{$\gamma=0.9$};
\addplot[
    color=color0, semithick, dashed,
    mark=triangle, mark repeat=40, mark options={solid}
    ]
    table[x=e,y=g9c] {fig/fig_data/bounds.dat};
\addlegendentry{$\gamma=0.9$ (CMAB)};
\addplot[
    color=color2, semithick,
    mark=*, mark repeat=40, mark phase=13
    ]
    table[x=e,y=g95] {fig/fig_data/bounds.dat};
\addlegendentry{$\gamma=0.95$};
\addplot[
    color=color2, semithick,dashed,
    mark=o, mark repeat=40, mark phase=13, mark options={solid}
    ]
    table[x=e,y=g95c] {fig/fig_data/bounds.dat};
\addlegendentry{$\gamma=0.95$ (CMAB)};
\addplot[
    color=color4, semithick,
    mark=square*, mark repeat=40, mark phase=27
    ]
    table[x=e,y=g99] {fig/fig_data/bounds.dat};
\addlegendentry{$\gamma=0.99$};
\addplot[
    color=color4, semithick, dashed,
    mark=square, mark repeat=40, mark phase=27, mark options={solid}
    ]
    table[x=e,y=g99c] {fig/fig_data/bounds.dat};
\addlegendentry{$\gamma=0.99$ (CMAB)};
\end{axis}

\def \zwidth {0.4\columnwidth}
\def \zheight {0.2\columnwidth}
\def \zwidth {0.4\columnwidth}
\def \zx {0} 
\def \zy {0.35\columnwidth} 

\begin{axis}[
    axis background/.style={fill=gray!10},
    at={(\zx,\zy)},
    width=\zwidth,
    height=\zheight,
    scale only axis,
    xmin=0, xmax=0.1,
    ymin=0, ymax=2,
    ymajorgrids=true,
    xmajorgrids=true,
    xtick={0,0.02,0.04,0.06,0.08,0.1},
    xticklabels={0,0.02,0.04,0.06,0.08,0.1},
    grid style={dashed,lightgray158},
    legend style={legend columns=1, font=\footnotesize, anchor=south, at={(0.5,0.02)}}
]

\addplot[
    color=color0, semithick,
    mark=triangle*, mark repeat=10
    ]
    table[x=e,y=g9] {fig/fig_data/bounds.dat};
\addplot[
    color=color0, semithick, dashed,
    mark=triangle, mark repeat=10, mark options={solid}
    ]
    table[x=e,y=g9c] {fig/fig_data/bounds.dat};
\addplot[
    color=color2, semithick,
    mark=*, mark repeat=10
    ]
    table[x=e,y=g95] {fig/fig_data/bounds.dat};
\addplot[
    color=color2, semithick, dashed,
    mark=o, mark repeat=10, mark options={solid}
    ]
    table[x=e,y=g95c] {fig/fig_data/bounds.dat};
\addplot[
    color=color4, semithick,
    mark=square*, mark repeat=10
    ]
    table[x=e,y=g99] {fig/fig_data/bounds.dat};
\addplot[
    color=color4, semithick, dashed,
    mark=square, mark repeat=10, mark options={solid}
    ]
    table[x=e,y=g99c] {fig/fig_data/bounds.dat};
\end{axis}

\end{tikzpicture}\vspace{-0.3cm}
\caption{Minimum value $\beta_{\min}$ that satisfies the condition for all \gls{ne} policies to be revealing as a function of the erasure probability $\varepsilon$.}
\label{fig:bound_general}
\end{figure}
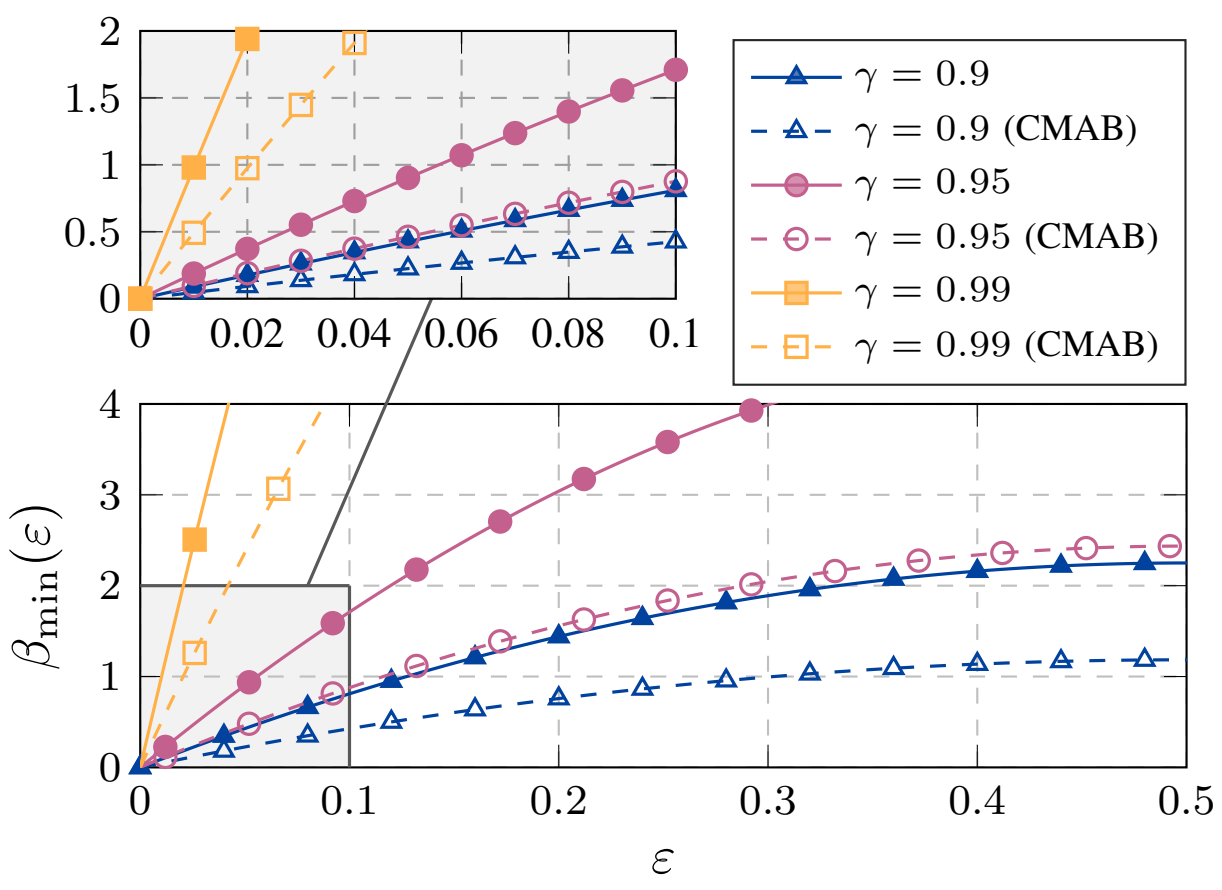

The minimum value $\beta_{\min}(\varepsilon)$ that satisfies~\eqref{eq:beta_cond} is shown in Fig.~\ref{fig:bound_general} for several practical values of the discount factor $\gamma$. We note that the values are shown for $\varepsilon\in[0,0.5]$, as the condition is symmetrical and $\beta_{\min}(\varepsilon)=\beta_{\min}(1-\varepsilon)$. 
Under realistic erasure probabilities $\varepsilon\in[0,0.1]$, the condition is not too restrictive for $\gamma=0.9$ and $\gamma=0.95$.
On the other hand, the condition quickly becomes very restrictive in the case with $\gamma=0.99$, requiring values of $\beta$ that will often lead the transmitter never to communicate at all.
This can be partially mitigated by substituting the $(1-\gamma)^{-1}$ factor with the highest possible long-term value in the selected \gls{pomdp} $V_{\max}$, which can be computed efficiently by setting $\beta=\varepsilon=\varphi=0$ and may be much lower.
Finally, we observe that the condition in~\eqref{eq:beta_cond} is sufficient, but may not be necessary, and, therefore, Corollary~\ref{cor:oneway_opt} may hold for a wider class of remote \glspl{pomdp}.

\subsection{Special Case: The Remote Contextual Bandit}
In the remainder of the section, we extend our analysis to remote \glspl{cmab}, which represent a subclass of remote \glspl{pomdp} in which the control actions do not affect the system evolution. Notably, remote monitoring applications, whose goal is only to track the state of the process and not to control it, are a special case of remote \glspl{cmab} in which $\mc{A}=\mc{S}$ and the reward function represents the quality of the estimate.
\begin{definition}
    A remote \gls{cmab} is a special case of remote \gls{pomdp} in which the receiver's actions do not affect future state transitions, i.e.,
    \begin{equation}
       P_{s,s'}^a=P_{s,s'}^{a'},\ \forall s,s'\in\mc{S},a,a'\in\mc{A}.
    \end{equation}
\end{definition}

\begin{theorem}\label{thm:cmab_time_inv}
  In the remote \gls{cmab} special case, if the joint policy $\bm{\pi}$ is an \gls{ne} and the problem respects condition
  \begin{equation}
    \beta>\gamma\varepsilon(1-\varepsilon)(1-\gamma^2)^{-1},\label{eq:beta_cmab_cond}
  \end{equation}
$\bm{\pi}$ must be revealing for any value of $\varphi$.
\end{theorem}
\begin{IEEEproof}
The proof is presented as Appendix B.
\end{IEEEproof}

In the remote \gls{cmab} scenario, the value of $\beta_{\min}$ from~\eqref{eq:beta_cond} can be divided by a factor $1+\gamma$ with respect to the general condition, which corresponds to almost halving the value of $\beta$ required for the optimality of revealing policies.
The value of the minimum communication cost $\beta_{\min}(\varepsilon)$ as a function of the forward channel erasure probability $\varepsilon$ is reported in Fig.~\ref{fig:bound_general}, with dashed lines indicating the values obtained when applying~\eqref{eq:beta_cmab_cond}. The condition is independent of the value of $\varphi$, as revealing policies completely obviate the need for an explicit feedback mechanism.
We finally observe that all our considerations on further improvements for specific remote \glspl{pomdp} and a wider applicability of Corollary~\ref{cor:oneway_opt} are valid in the remote \gls{cmab} scenario as well.

\section{Simulation Settings and Results}\label{sec:simulation}

In this section, we verify the correctness of our theoretical analysis and examine how conservative the sufficient conditions for optimality are through Monte Carlo simulations in a family of remote \glspl{pomdp}.

We first note that the performance of the optimal policy $\bm{\pi}_{\varphi=0}^{*}$ with ideal feedback represents an upper bound to the performance of the system.
Lemma~\ref{lm:phi0-optimality} ensures that, if the optimal policy is revealing, it will also be playable for any $\varphi>0$, making the upper bound valid for all such conditions.
Notably, if the optimal policy in the $\varphi=0$ case is not revealing and the condition in Theorem~\ref{thm:oneway_cond} is not verified, we do not have an efficient strategy to find an optimal policy for $\varphi>0$.

We then consider an implementable scheme, which provides a lower bound for optimal performance. We consider a case in which the two nodes continue using $\bm{\pi}_{\varphi=0}^{*}$, but, whenever the transmitter sends a message $m_t$ for which the receiver policy is non-revealing at the current history $h_{\text{rx}}$, the receiver applies the action prescribed by the original policy, but updates its effective history with $\chi$ instead of $s$. In other words, the transmitter ignores messages that could lead to ambiguity, and since the transmitter knows this rule, it can track the same effective history. The two nodes therefore remain synchronized, even though the action does not reveal the channel outcome. We make no claims as to the optimality of this strategy, but note that it represents a simple lower bound to optimal performance.

\subsection{MDP Generation}

We consider a family of \glspl{mdp} with $S=6$ states and $A=2$ control actions to drive the evolution of the system.
For each state and action, we then define a target state $\zeta(s,a)$ such that we have the following transition probability function:
\begin{equation}
    P_{s,s'}^a=\begin{cases}
        \frac{9}{14},&\text{if }s'=\zeta(s,a);\\
        \frac{5}{28},&\text{if }|s'-\zeta(s,a)|\in\{1,S-1\}.
    \end{cases}
\end{equation}
In this way, each action has $3$ possible outcomes, which correspond to the target state $\zeta(s,a)$ and its neighbors when considering circular coordinates.
In all cases, the transition matrices are checked to ensure that the \gls{mdp} is recurrent, i.e., each state is reachable from any other state with a non-zero probability under a random policy.
The receiver reward is defined as $r(s,a,s')=\exp\left(-\lambda |s'-s^\star|\right)$, where $s^\star$ is a randomly picked state. By randomly selecting the parameters $\zeta$ and $s^\star$, we thus generate a family of random \glspl{mdp} with similar characteristics, corresponding to the one studied in~\cite{talli2025pragmatic} for the ideal communication case.

In order to examine the \gls{cmab} case, we maintain the same transition probability matrix and examine two separate cases. 
Firstly, we consider a remote estimation task, in which $\mc{A}=\mc{S}$ and $r(s,a,s')$ is $1$ if $a=s$ and $0$ otherwise.
Secondly, we define a \gls{cmab} with two actions $\mathcal{A}=\{0,1\}$, defining $r(s,a,\cdot)$ $\forall$ $a \in \mathcal{A}$ by sampling a uniform distribution $\mc{U}(0,1)$.
In both cases, actions do not affect state transitions.

\subsection{Revealing Policy Conditions}

We first consider the optimal policy resulting from \gls{jpo} in the ideal case with $\varphi=0$: following Lemma~\ref{lm:phi0-optimality}, if $\bm{\pi}^*_{\varphi=0}$ revealing, it will remain optimal for any $\varphi>0$. We thus analyze the class $\bm{\pi}^*_{\varphi=0}$ belongs to over an \gls{mdp} defined as above, checking whether the revealing condition holds.
Naturally, this is not representative of all possible \glspl{mdp}, but it can provide a qualitative idea of the conservativeness of Theorem~\ref{thm:oneway_cond} and the optimality of revealing policies.

\begin{figure}[t]
\centering
\input{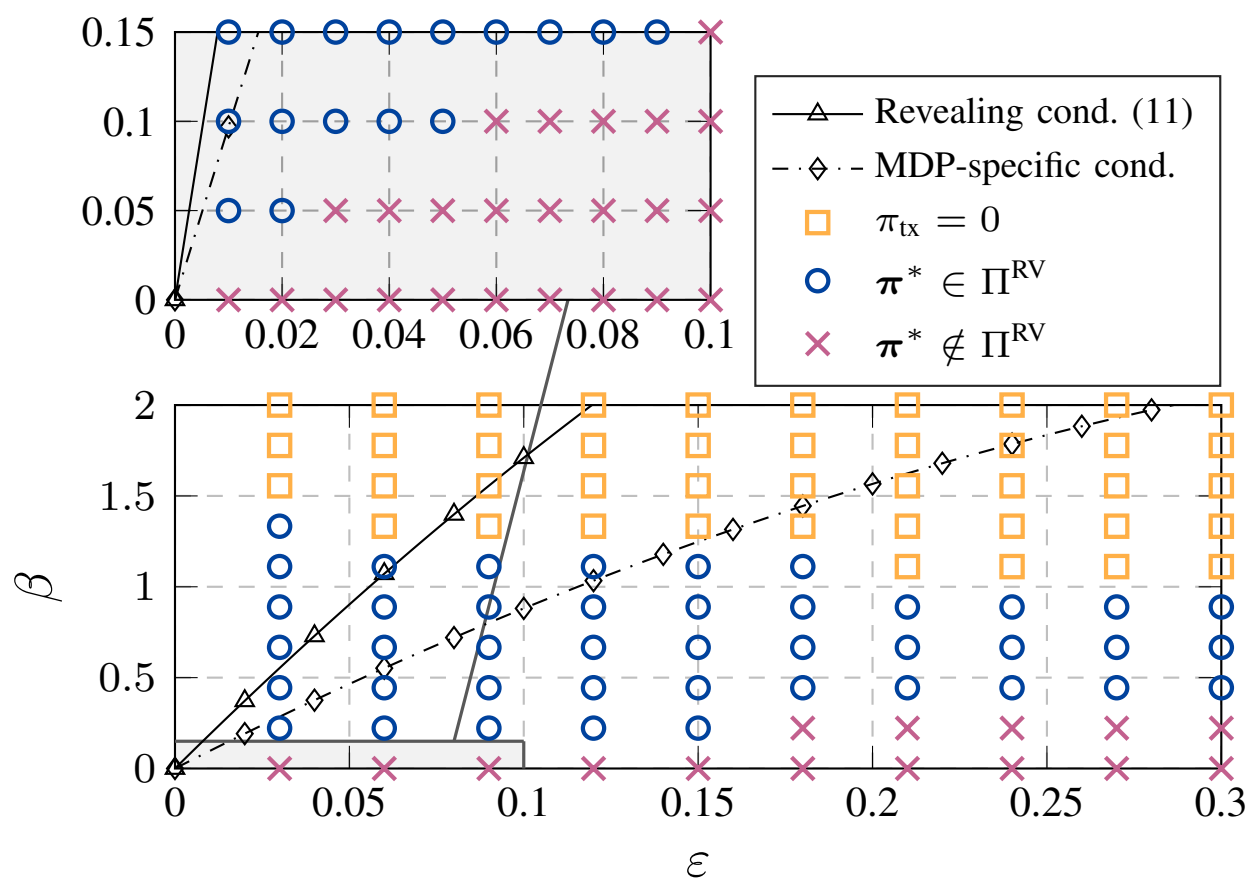}\vspace{-0.3cm}
\caption{Analysis of the conditions for revealing policy optimality in a specific remote \gls{pomdp} with $\gamma=0.95$.}\vspace{-0.3cm}
\label{fig:control}
\end{figure}

Fig.~\ref{fig:control} shows the nature of $\bm{\pi}^*_{\varphi=0}$, obtained through \gls{jpo}, for different values of $\beta$ and $\varepsilon$, setting $\gamma=0.95$. Theorem~\ref{thm:oneway_cond} is quite conservative, as non-revealing policies appear only for very low values of $\beta$, as the zoomed-in plot show, and so is the tighter bound obtained by computing the highest possible long-term reward $V_{\max}$ of the \gls{mdp} with ideal communication. We also observe that the optimal policy is never to transmit if $\beta>1$ and $\varepsilon$ is sufficiently high: this policy is still revealing, albeit with degenerate properties.

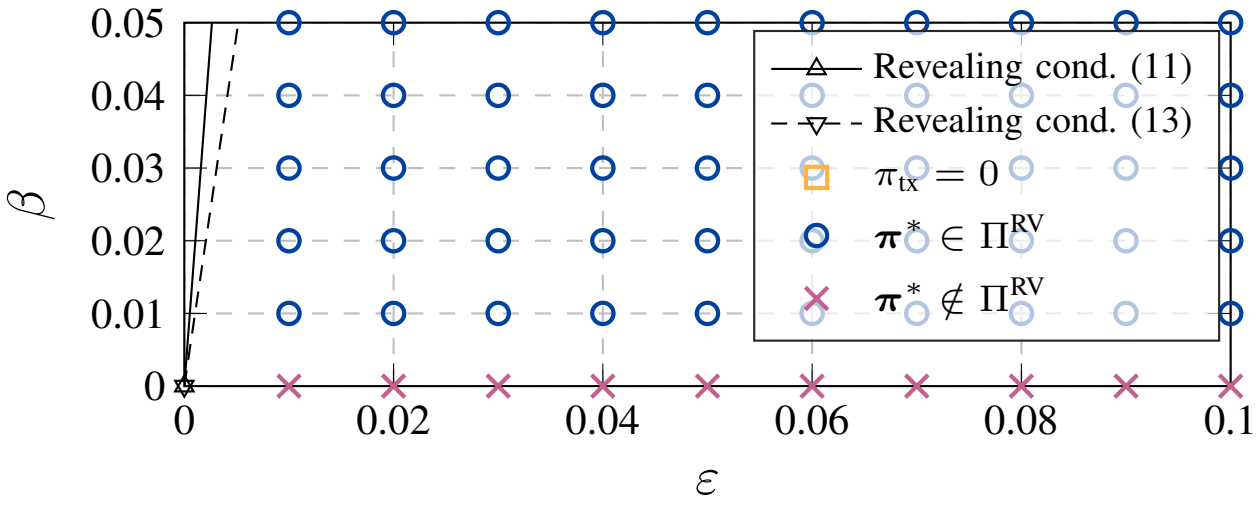
\begin{figure}[t]
\centering
\begin{tikzpicture}
\pgfplotsset{scaled y ticks=false}
\begin{axis}[
    clip mode=individual,
    at={(0,0)},
    width=\fwidth,
    height=\fheight,
    xlabel={$\varepsilon$},
    ylabel={$\beta$},
    xmin=0, xmax=0.1,
    xtick={0,0.02,0.04,0.06,0.08,0.1},
    xticklabels={0,0.02,0.04,0.06,0.08,0.1},
    ymin=0, ymax=0.05,
    ytick={0,0.01,0.02,0.03,0.04,0.05},
    yticklabels={0,0.01,0.02,0.03,0.04,0.05},
    legend pos=south east,
    ymajorgrids=true,
    xmajorgrids=true,
    grid style=dashed,
    legend style={legend columns=1, font=\scriptsize, anchor=north east, at={(0.99,0.98)}}
]

\addplot[
    color=black, mark=triangle,mark repeat=10
    ]
    table{
0	0
0.001	0.018981
0.002	0.037924
0.003	0.056829
0.004	0.0756959999999999
0.005	0.0945249999999999
0.006	0.113316
0.007	0.132069
0.008	0.150784
0.009	0.169461
0.01	0.1881
0.011	0.206701
0.012	0.225264
0.013	0.243789
0.014	0.262276
0.015	0.280725
0.016	0.299136
0.017	0.317509
0.018	0.335844
0.019	0.354141
0.02	0.3724
0.021	0.390621
0.022	0.408804
0.023	0.426949
0.024	0.445056
0.025	0.463125
0.026	0.481156
0.027	0.499149
0.028	0.517104
0.029	0.535021
0.03	0.552899999999999
0.031	0.570741
0.032	0.588544
0.033	0.606308999999999
0.034	0.624036
0.035	0.641725
0.036	0.659375999999999
0.037	0.676988999999999
0.038	0.694563999999999
0.039	0.712100999999999
0.04	0.729599999999999
0.041	0.747060999999999
0.042	0.764483999999999
0.043	0.781868999999999
0.044	0.799215999999999
0.045	0.816524999999999
0.046	0.833795999999999
0.047	0.851028999999999
0.048	0.868223999999999
0.049	0.885380999999999
0.05	0.902499999999999
0.051	0.919580999999999
0.052	0.936623999999999
0.053	0.953628999999999
0.054	0.970595999999999
0.055	0.987524999999999
0.056	1.004416
0.057	1.021269
0.058	1.038084
0.059	1.054861
0.06	1.0716
0.061	1.088301
0.062	1.104964
0.063	1.121589
0.064	1.138176
0.065	1.154725
0.066	1.171236
0.067	1.187709
0.068	1.204144
0.069	1.220541
0.07	1.2369
0.071	1.253221
0.072	1.269504
0.073	1.285749
0.074	1.301956
0.075	1.318125
0.076	1.334256
0.077	1.350349
0.078	1.366404
0.079	1.382421
0.08	1.3984
0.081	1.414341
0.082	1.430244
0.083	1.446109
0.084	1.461936
0.085	1.477725
0.086	1.493476
0.087	1.509189
0.088	1.524864
0.089	1.540501
0.09	1.5561
0.091	1.571661
0.092	1.587184
0.093	1.602669
0.094	1.618116
0.095	1.633525
0.096	1.648896
0.097	1.664229
0.098	1.679524
0.099	1.694781
0.1	1.71
};
\addlegendentry{Revealing cond.~\eqref{eq:beta_cond}}

\addplot[color=black,densely dashed, mark=triangle,mark repeat=10,mark options={solid,rotate=180}]
    table{
0	0
0.001	0.00973384615384615
0.002	0.0194482051282051
0.003	0.0291430769230769
0.004	0.0388184615384615
0.005	0.0484743589743589
0.006	0.0581107692307692
0.007	0.0677276923076923
0.008	0.0773251282051281
0.009	0.0869030769230769
0.01	0.0964615384615384
0.011	0.106000512820513
0.012	0.11552
0.013	0.12502
0.014	0.134500512820513
0.015	0.143961538461538
0.016	0.153403076923077
0.017	0.162825128205128
0.018	0.172227692307692
0.019	0.181610769230769
0.02	0.190974358974359
0.021	0.200318461538461
0.022	0.209643076923077
0.023	0.218948205128205
0.024	0.228233846153846
0.025	0.2375
0.026	0.246746666666666
0.027	0.255973846153846
0.028	0.265181538461538
0.029	0.274369743589743
0.03	0.283538461538461
0.031	0.292687692307692
0.032	0.301817435897436
0.033	0.310927692307692
0.034	0.320018461538461
0.035	0.329089743589743
0.036	0.338141538461538
0.037	0.347173846153846
0.038	0.356186666666666
0.039	0.36518
0.04	0.374153846153846
0.041	0.383108205128205
0.042	0.392043076923077
0.043	0.400958461538461
0.044	0.409854358974359
0.045	0.418730769230769
0.046	0.427587692307692
0.047	0.436425128205128
0.048	0.445243076923077
0.049	0.454041538461538
0.05	0.462820512820512
0.051	0.47158
0.052	0.48032
0.053	0.489040512820512
0.054	0.497741538461538
0.055	0.506423076923076
0.056	0.515085128205128
0.057	0.523727692307692
0.058	0.532350769230769
0.059	0.540954358974359
0.06	0.549538461538461
0.061	0.558103076923077
0.062	0.566648205128205
0.063	0.575173846153846
0.064	0.58368
0.065	0.592166666666666
0.066	0.600633846153846
0.067	0.609081538461538
0.068	0.617509743589743
0.069	0.625918461538461
0.07	0.634307692307692
0.071	0.642677435897435
0.072	0.651027692307692
0.073	0.659358461538461
0.074	0.667669743589743
0.075	0.675961538461538
0.076	0.684233846153846
0.077	0.692486666666666
0.078	0.70072
0.079	0.708933846153846
0.08	0.717128205128205
0.081	0.725303076923076
0.082	0.733458461538461
0.083	0.741594358974358
0.084	0.749710769230769
0.085	0.757807692307692
0.086	0.765885128205128
0.087	0.773943076923076
0.088	0.781981538461538
0.089	0.790000512820512
0.09	0.797999999999999
0.091	0.805979999999999
0.092	0.813940512820512
0.093	0.821881538461538
0.094	0.829803076923076
0.095	0.837705128205127
0.096	0.845587692307692
0.097	0.853450769230769
0.098	0.861294358974358
0.099	0.869118461538461
0.1	0.876923076923076
};
\addlegendentry{Revealing cond.~\eqref{eq:beta_cmab_cond}}

\addplot[
    color=color4, thick,
    mark=square, only marks
    ]
    table{
    -0.01   0.01
    }
    ;
\addlegendentry{$\pi_{\text{tx}}=0$};

\addplot[
    color=color0, thick,
    mark=o, only marks
    ]
    table[x=eps,y=beta] {fig/fig_data/cmab/remote_estimation/0p95/est_zoom_rev_0p95.dat};
\addlegendentry{$\bm{\pi}^*\in\Pi^{\text{RV}}$};
\addplot[
    color=color2, thick,
    mark=x, only marks,mark size=3
    ]
    table[x=eps,y=beta] {fig/fig_data/cmab/remote_estimation/0p95/est_zoom_nonrev_0p95.dat};
\addlegendentry{$\bm{\pi}^*\notin\Pi^{\text{RV}}$};
\end{axis}
\end{tikzpicture}\vspace{-0.3cm}
\caption{Analysis of the conditions for revealing policy optimality in a specific remote monitoring problem with $\gamma=0.95$.}
\label{fig:estimation}
\end{figure}

\begin{figure}[t]
\centering
\input{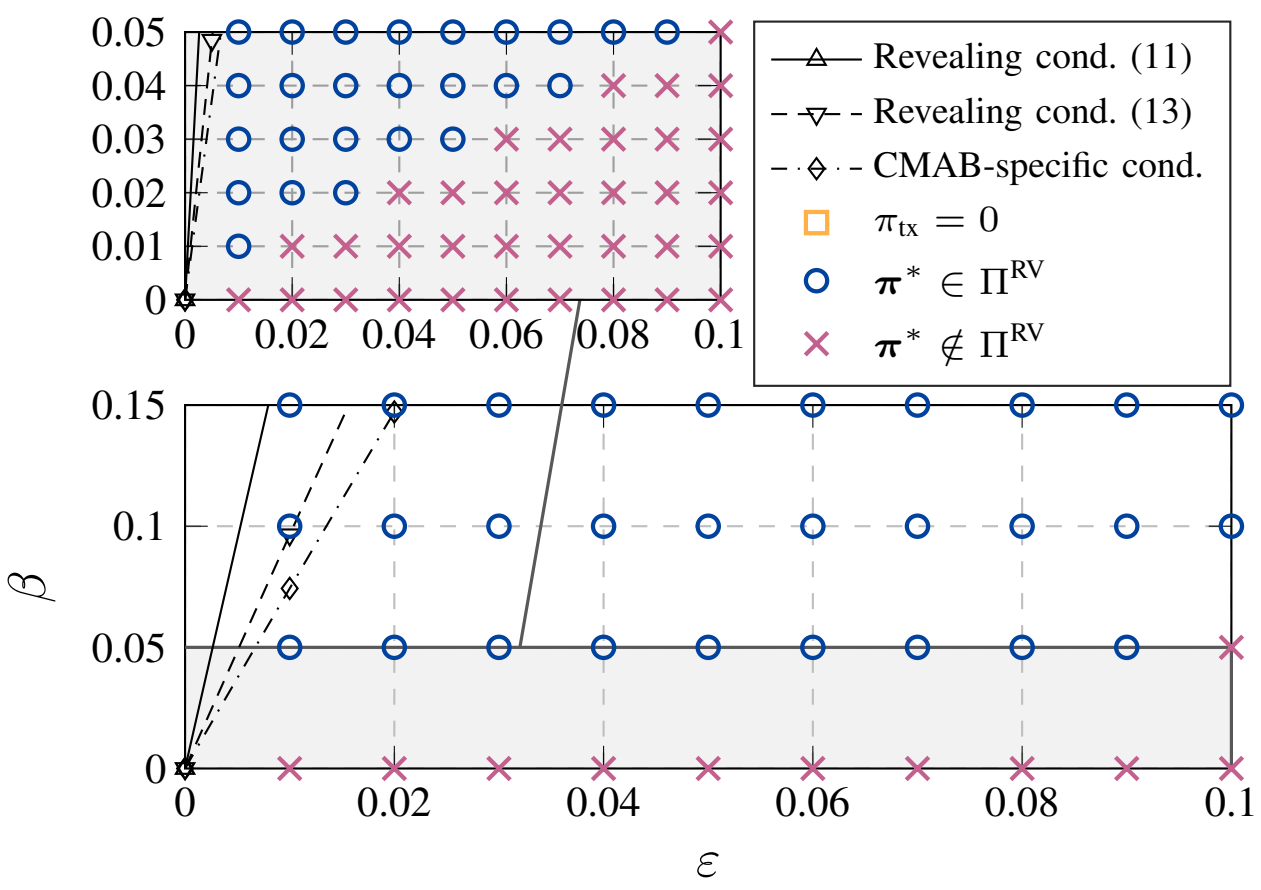}\vspace{-0.3cm}
\caption{Analysis of the conditions for revealing policy optimality in a specific two-action remote \gls{cmab} with $\gamma=0.95$.}
\label{fig:cmab}
\end{figure}

The region of revealing policies is even larger in the remote estimation case, shown in Fig.~\ref{fig:estimation}, where we obtain non-revealing policies only for $\beta=0$: in this scenario, the larger number of actions makes it easier for the transmitter to distinguish between different states.
We have a few more non-revealing policies in the two-action \gls{cmab} case, shown in Fig.~\ref{fig:cmab}, but the region in which $\bm{\pi}^*_{\varphi=0}$ is revealing is significantly larger than in the control scenario.
Therefore, the condition for the \gls{cmab} case is also extremely conservative, and the applicability of \gls{ack}-free \gls{goc} strategies is thus much wider than the sufficient theoretical condition would suggest.

\subsection{Optimality Performance Gap}

The results we presented above confirm that there is a wide set of parameters $\varepsilon$ and $\beta$ for which policies that are optimal under perfect feedback are revealing. However, we also investigate settings in which this is not the case, using performance with an ideal \gls{ack} channel as an upper bound and the lower bound scheme we described above. By evaluating the gap between these bounds, we can know the potential gain of optimizing complex history-dependent strategies. 

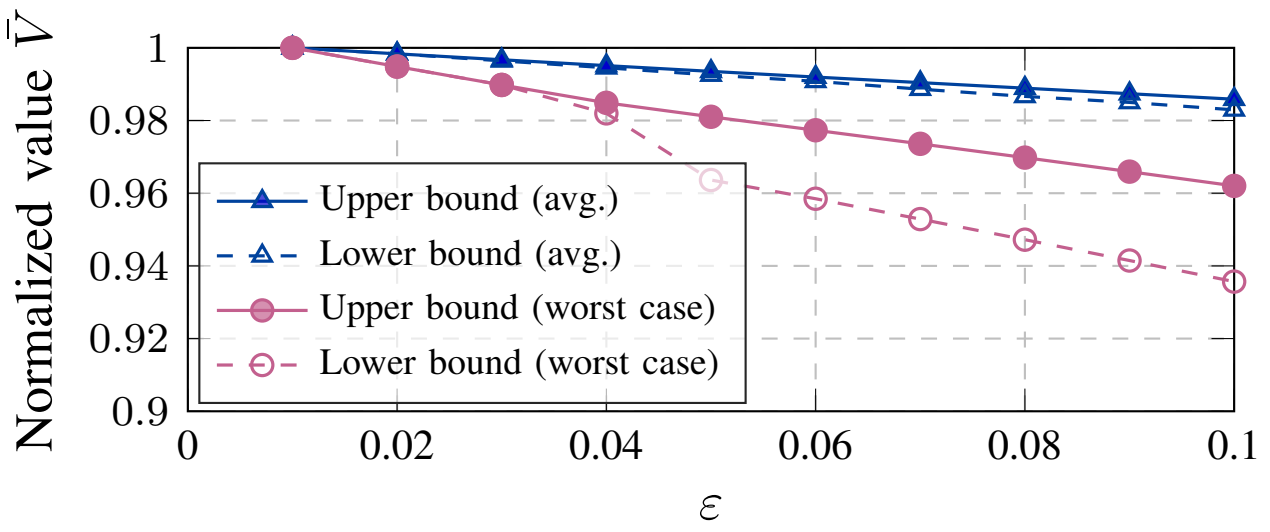
\begin{figure}[t]
\centering
\begin{tikzpicture}
\begin{axis}[
    clip mode=individual,
    at={(0,0)},
    width=\fwidth,
    height=\fheight,
    xlabel={$\varepsilon$},
    ylabel={Normalized value $\bar{V}$},
    xmin=0, xmax=0.1,
    domain=0:0.1,
    ymin=0.9, ymax=1,
    legend pos=south east,
    ymajorgrids=true,
    xmajorgrids=true,
    grid style=dashed,
    xtick={0,0.02,0.04,0.06,0.08,0.1},
    xticklabels={0,0.02,0.04,0.06,0.08,0.1},
    legend style={legend columns=1, font=\scriptsize, anchor=south west, at={(0.01,0.02)}}
]

\addplot+[color=color0, semithick,
    mark=triangle*,name path=upavg]
    table[x=epsilon,y=upper_avg] {fig/fig_data/control/0p95/control_gap_0p95.dat};
\addlegendentry{Upper bound (avg.)};
\addplot+[color=color0, semithick,dashed,
    mark=triangle, mark options={solid},name path=lowavg
    ]
    table[x=epsilon,y=lower_avg] {fig/fig_data/control/0p95/control_gap_0p95.dat};
\addlegendentry{Lower bound (avg.)};

\addplot[name path=upworst,
    color=color2, semithick,
    mark=*
    ]
    table[x=epsilon,y=upper_maxgap] {fig/fig_data/control/0p95/control_gap_0p95.dat};
\addlegendentry{Upper bound (worst case)};
\addplot[name path=lowworst,
    color=color2, semithick,dashed,
    mark=o, mark options={solid}
    ]
    table[x=epsilon,y=lower_maxgap] {fig/fig_data/control/0p95/control_gap_0p95.dat};
\addlegendentry{Lower bound (worst case)};

\end{axis}
\end{tikzpicture}\vspace{-0.3cm}
\caption{Analysis of the optimality gap over $60$ instances of remote \glspl{pomdp} with $\gamma=0.95$ and $\beta=0.05$.}
\label{fig:control_gap}
\end{figure}

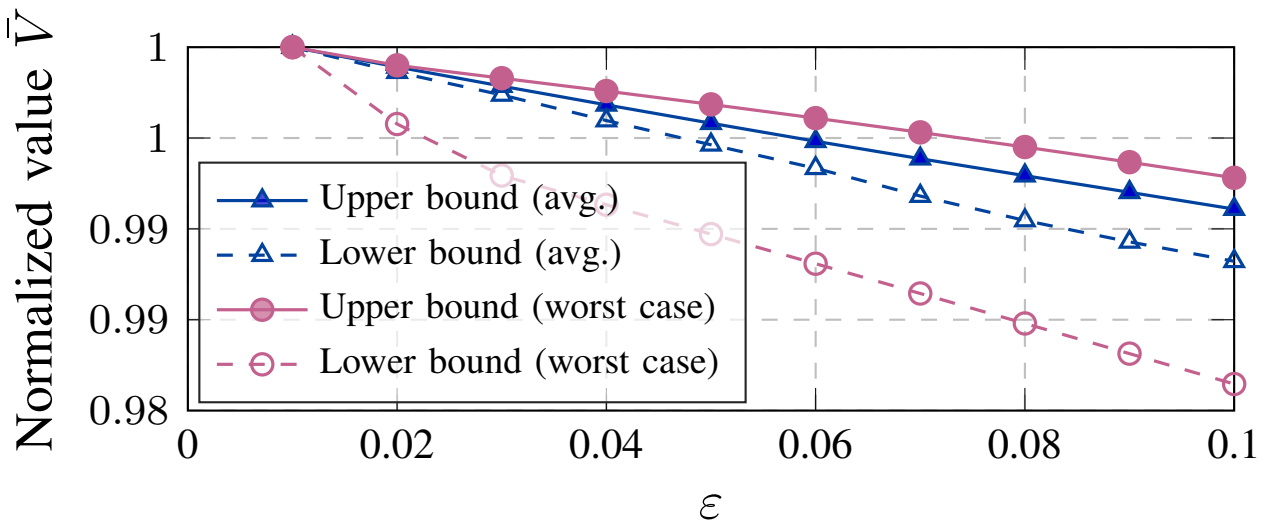
\begin{figure}[t]
\centering
\begin{tikzpicture}
\begin{axis}[
    clip mode=individual,
    at={(0,0)},
    width=\fwidth,
    height=\fheight,
    xlabel={$\varepsilon$},
    ylabel={Normalized value $\bar{V}$},
    xmin=0, xmax=0.1,
    domain=0:0.1,
    ymin=0.98, ymax=1,
    legend pos=south east,
    ymajorgrids=true,
    xmajorgrids=true,
    grid style=dashed,
    xtick={0,0.02,0.04,0.06,0.08,0.1},
    xticklabels={0,0.02,0.04,0.06,0.08,0.1},
    legend style={legend columns=1, font=\scriptsize, anchor=south west, at={(0.01,0.02)}}
]

\addplot+[color=color0, semithick,
    mark=triangle*,name path=upavg]
    table[x=epsilon,y=upper_avg] {fig/fig_data/cmab/random_reward/0p95/rr_gap_0p95.dat};
\addlegendentry{Upper bound (avg.)};
\addplot+[color=color0, semithick,dashed,
    mark=triangle, mark options={solid},name path=lowavg
    ]
    table[x=epsilon,y=lower_avg] {fig/fig_data/cmab/random_reward/0p95/rr_gap_0p95.dat};
\addlegendentry{Lower bound (avg.)};

\addplot[name path=upworst,
    color=color2, semithick,
    mark=*
    ]
    table[x=epsilon,y=upper_maxgap] {fig/fig_data/cmab/random_reward/0p95/rr_gap_0p95.dat};
\addlegendentry{Upper bound (worst case)};
\addplot[name path=lowworst,
    color=color2, semithick,dashed,
    mark=o, mark options={solid}
    ]
    table[x=epsilon,y=lower_maxgap] {fig/fig_data/cmab/random_reward/0p95/rr_gap_0p95.dat};
\addlegendentry{Lower bound (worst case)};

\end{axis}
\end{tikzpicture}\vspace{-0.3cm}
\caption{Analysis of the optimality gap over $60$ instances of CMAB random reward with $\gamma=0.95$ and $\beta=0.02$.}
\label{fig:cmab_gap}
\end{figure}

We fix $\gamma=0.95$ and consider two possible communication costs: $\beta=0.05$ for the control problem, and $\beta=0.02$ for the \gls{cmab} problem.
Hence, we vary $\varepsilon \in [0, 0.1]$ and compute the performance of $\bm{\pi}^*_{\varphi=0}$ and the lower bound scheme for each configuration, considering $60$ random  \gls{mdp} instances from the same family.
The results are shown in Fig.~\ref{fig:control_gap} for the control scenario. We can see that, even in the worst case, i.e., the \gls{mdp} with the widest gap, the performance loss is below $3\%$ for $\varepsilon\leq0.1$. In $75\%$ of cases, the two are actually identical, i.e., $\bm{\pi}^*_{\varphi=0}$ is revealing even under these settings. In the \gls{cmab} case, shown in Fig.~\ref{fig:cmab_gap}, about two thirds of cases have a non-zero optimality gap, but even in the worst case the gap is close to $1\%$. This corroborates our view that revealing policies represent an effective solution to the imperfect feedback problem, allowing for time-invariant representations without affecting performance.

\section{Conclusion and Future Work}\label{sec:conc}

In this work, we presented a theoretical analysis of remote \glspl{pomdp} with unreliable communication, considering scenarios with an imperfect \gls{ack} channel.
We introduced the notion of \emph{revealing} policies, which allow the system agents to preserve an information hierarchy even without relying on explicit \glspl{ack}, and studied their optimality.
We showed that revealing policies, which always have a time-invariant equivalent, are optimal for a wide range of system parameters, even wider for remote contextual bandits, deriving analytical bounds and discussing their conservative nature. 

Our theoretical results pave the way for new approaches for optimizing remote \glspl{pomdp}, and this study can be extended in several ways: on the practical side, applying these results to modern remote reinforcement learning schemes would be extremely interesting.
On the theoretical side, the analysis could extend to settings where the receiver can distinguish silence from channel erasures, which introduces an additional side channel between the agents.

\section*{Use of AI Disclosure}

The AI system GPT-5.6 Sol was used to implement parts of the simulation code for the results in Sec.~\ref{sec:simulation}. The code was then checked for correctness by the authors and will be published upon acceptance for reproducibility.

\appendix
\section*{A. Proof of Theorem~\ref{thm:oneway_cond}}
We prove the Theorem by induction: if its statement is true up to step $t-1$,
Lemma~\ref{lm:implicit_sync} allows us to state that any possible transmitter-side history $h_{\text{tx}}$ up to time $t$ is synchronized. If there is at least one state $s$ in which the transmitter sends an update but the receiver policy is not revealing, we define a new transmitter policy $\pi_{\text{tx}}'$, and show that $V^{\pi_{\text{tx}}',\pi_{\text{rx}}}_{s}(h_{\text{tx}},h_{\text{rx}})>V^{\pi_{\text{tx}},\pi_{\text{rx}}}_{s}(h_{\text{tx}},h_{\text{rx}})$. The new policy is
\begin{equation}
  \pi'_{\text{tx},\tau}(h_{\text{tx}}')=
  \begin{cases}
    \pi_{\text{tx},\tau}(h_{\text{tx}}'), &\tau<t\vee h_{\text{tx}}\not\subset h_{\text{tx}}';\\
    0, &\tau=t;\\
    1, &\tau=t+1;\\
    \pi_{\text{tx},\tau}(h_{\text{tx}}''), &\text{otherwise,}
  \end{cases}
\end{equation}
where $h_{\text{tx}}''$ is a fictitious history which is identical to $h_{\text{tx}}'$ except for step $t$, in which $c''_t=1$. This ensures that the two policies behave identically from step $t+2$ onward.
 In other words, the modified policy removes the transmission attempt at time $t$, forcing a transmission at time $t+1$. This is equivalent to removing a redundant transmission if the original policy would have transmitted in step $t+1$, and to delaying the transmission by one step otherwise. We then decompose $V^{\pi_{\text{tx}},\pi_{\text{rx}}}_{s,t}(h_{\text{tx}},h_{\text{rx}})$:
\begin{equation}\label{eq:orig_val_onestep}
\begin{aligned}
  V^{\pi_{\text{tx}},\pi_{\text{rx}}}_{s}(h_{\text{tx}},h_{\text{rx}})=-\beta+\sum_{s'\in\mc{S}}P_{s,s'}^a\big[r(s,a,s')\\
  +\gamma V^{\pi_{\text{tx}},\pi_{\text{rx}}}_{s'}([h_{\text{tx}},\langle1,a,f_t,s'\rangle],[h_{\text{rx}},\langle m_t,a\rangle])\big].
\end{aligned}
\end{equation}
where $a=\pi_{\text{rx},t}(h_{\text{rx}},\chi)=\pi_{\text{rx},t}(h_{\text{rx}},s)$ is the receiver's action at time $t$. 
We now define utility variable $o_{\text{tx}}^{f,s'}=\langle1,a,f,s'\rangle$.
We can then give the probability $p_{\text{tx}}(s')$ that the transmitter will send another update in step $t+1$ after transitioning to state $s'$, given that the message at time $t$ was correctly received, as
\begin{equation}
  p_{\text{tx}}(s')=(1-\varphi)\pi_{\text{tx},t+1}\!\left(\left[h_{\text{tx}},o_{\text{tx}}^{1,s'}\right]\right)+\varphi\pi_{\text{tx},t+1}\!\left(\left[h_{\text{tx}},o_{\text{tx}}^{0,s'}\right]\right)\!.
\end{equation}
Let us now consider the value improvement $\Delta V$, defined as
\begin{equation}
  \Delta V=V_{s}^{\pi_{\text{tx}}',\pi_{\text{rx}}}(h_{\text{tx}},h_{\text{rx}})-V_{s}^{\pi_{\text{tx}},\pi_{\text{rx}}}(h_{\text{tx}},h_{\text{rx}}).
\end{equation}
We first note that the two only differ if the message at time $t$ is correctly receive, aside from the additional cost $\beta$ which is incurred by the original policy $\pi_{\text{tx}}$ in any case.

We define a \emph{post-decision state}~\cite{salodkar2008online}, which describes the system after the transmitter's action but before the receiver's.
The post-decision state at any time step is characterized by tuple $\langle s,c,m\rangle$: naturally, if $c=0$, the message received is $m=\chi$ with probability $1$.
The value of the post-decision state $\langle s,c,m\rangle$ at time $t$ is
\begin{equation}\label{eq:value_rx}
\begin{aligned}
  U_{s,c,m}^{\pi_{\text{tx}},\pi_{\text{rx}}}(h_{\text{tx}},h_{\text{rx}})=\sum_{a\in\mc{A}}\pi_{\text{rx},t}(a|h_{\text{rx}},m)\sum_{s'\in\mc{S}}P_{s,s'}^{a}\big[r(s,a,s')
  \\+\gamma V_{s'}^{\pi_{\text{tx}},\pi_{\text{rx}}}([h_{\text{tx}},\langle c,a,s'\rangle],[h_{\text{rx}},\langle m,a\rangle])\big].
\end{aligned}
\end{equation}
We can compute the value improvement by conditioning on the outcome of $m_t$ and considering post-decision states, obtaining
\begin{equation}\label{eq:delta_val}
\begin{aligned}
  \Delta V=\beta+\gamma(1-\varepsilon)\sum_{\mathclap{s'\in\mc{S}}}\Bigg[P_{s,s'}^a p_{\text{tx}}(s')\Bigg((1-\varepsilon)\\ \times\bigg[U^{\pi_{\text{tx}},\pi_{\text{rx}}}_{s',1,s'}([h_{\text{tx}},o_{\text{tx}}^{0,s'}],h_{\text{rx}}^{(\chi)})- U^{\pi_{\text{tx}},\pi_{\text{rx}}}_{s',1,s'}([h_{\text{tx}},o_{\text{tx}}^{f_t,s'}],h_{\text{rx}}^{(s)})\bigg]\\
  +\varepsilon\bigg[U^{\pi_{\text{tx}},\pi_{\text{rx}}}_{s',1,\chi}([h_{\text{tx}},o_{\text{tx}}^{0,s'}],h_{\text{rx}}^{(\chi)})  -U^{\pi_{\text{tx}},\pi_{\text{rx}}}_{s',1,\chi}([h_{\text{tx}},o_{\text{tx}}^{f_t,s'}],h_{\text{rx}}^{(s)})
  \bigg]\Bigg)\\
  +P_{s,s'}^a (1-p_{\text{tx}}(s'))\bigg[\big(U^{\pi_{\text{tx}},\pi_{\text{rx}}}_{s',1,s'}([h_{\text{tx}},o^{0,s'}_{\text{tx}}],h_{\text{rx}}^{(\chi)})\\
  -U^{\pi_{\text{tx}},\pi_{\text{rx}}}_{s',0,\chi}([h_{\text{tx}},o^{f_t,s'}_{\text{tx}}],h_{\text{rx}}^{(s)})\big)-\beta\bigg]\Bigg],
\end{aligned}
\end{equation}
where $h_{\text{rx}}^{(s)}=[h_{\text{rx}},\langle s,a\rangle]$ and $h_{\text{rx}}^{(\chi)}=[h_{\text{rx}},\langle \chi,a\rangle]$.
We first note that, due to the Markov property, receiving a message at time $t+1$ cannot decrease future value with respect to receiving it at time $t$, as the receiver has better information about the state: accordingly, policy $\pi'_{\text{tx}}$ always improves the long-term value if the original policy does not transmit in step $t+1$, and we get condition
\begin{equation}
\begin{aligned}
\Delta V=\beta+\gamma(1-\varepsilon)\sum_{\mathclap{s'\in\mc{S}}}P_{s,s'}^a p_{\text{tx}}(s')\Bigg((1-\varepsilon)\\ \times\bigg[U^{\pi_{\text{tx}},\pi_{\text{rx}}}_{s',1,s'}([h_{\text{tx}},o_{\text{tx}}^{0,s'}],h_{\text{rx}}^{(\chi)})- U^{\pi_{\text{tx}},\pi_{\text{rx}}}_{s',1,s'}([h_{\text{tx}},o_{\text{tx}}^{f_t,s'}],h_{\text{rx}}^{(s)})\bigg]\\
  +\varepsilon\bigg[U^{\pi_{\text{tx}},\pi_{\text{rx}}}_{s',1,\chi}([h_{\text{tx}},o_{\text{tx}}^{0,s'}],h_{\text{rx}}^{(\chi)})  -U^{\pi_{\text{tx}},\pi_{\text{rx}}}_{s',1,\chi}([h_{\text{tx}},o_{\text{tx}}^{f_t,s'}],h_{\text{rx}}^{(s)})
  \bigg]\Bigg).
\end{aligned}
\end{equation}
Thanks to the Markov property, receiving both $s_t$ and $s_{t+1}$ provides the receiver with the same information as only receiving $s_{t+1}$. If $\pi_{\text{tx}}$ is the best response to $\pi_{\text{rx}}$, the first two terms of the sum must thus be identical, and we obtain
\begin{equation}
\begin{aligned}
  \Delta V\geq\beta+\gamma\sum_{s'\in\mc{S}
  }P_{s,s'}^a p_{\text{tx}}(s')\varepsilon(1-\varepsilon)\\
  \times\left[U^{\pi_{\text{tx}},\pi_{\text{rx}}}_{s',1,\chi}([h_{\text{tx}},o_{\text{tx}}^{0,s'}],h_{\text{rx}}^{(\chi)})  -U^{\pi_{\text{tx}},\pi_{\text{rx}}}_{s',1,\chi}([h_{\text{tx}},o_{\text{tx}}^{f_t,s'}],h_{\text{rx}}^{(s)})
  \right].
\end{aligned}
\end{equation}
We denote the minimum and maximum possible long-term rewards for the \gls{mdp} with ideal communication by $V_{\min}$ and $V_{\max}$, respectively. Since the instantaneous reward is in interval $[0,1]$, we know that $V_{\min}\geq0$ and $V_{\max}\leq(1-\gamma)^{-1}$. We thus compute a more conservative condition that holds for any \gls{mdp} definition:
\begin{equation}
\begin{aligned}
  \Delta V\geq\beta-\gamma\sum_{s'\in\mc{S}} \varepsilon(1-\varepsilon)P_{s,s'}^a p_{\text{tx}}(s')(1-\gamma)^{-1}.
\end{aligned}
\end{equation}
As $\sum_{s'\in\mc{S}_{\text{tx}}}P_{s,s'}^a p_{\text{tx}}(s')\leq1$, we obtain
\begin{equation}
\begin{aligned}
  \Delta V\geq\beta-\gamma\varepsilon(1-\varepsilon)(1-\gamma)^{-1}.
\end{aligned}
\end{equation}
It is thus easy to see that, under the condition in~\eqref{eq:beta_cond}, the value difference is positive, and $\pi_{\text{tx}}$ is not the best response to $\pi_{\text{rx}}$. We can repeat this for all values of $s$ in which $\pi_{\text{rx}}$ is not revealing, obtaining a new transmission policy that maintains synchronization at step $t$. We can thus extend the result to any time step $t$.

\section*{B. Proof of Theorem~\ref{thm:cmab_time_inv}}

As we did in the proof of Theorem~\ref{thm:oneway_cond}, we consider a policy such that any history $h_{\text{tx}}$ is synchronized up to time $t$, and a state $s_t=s$ in which $\pi_{\text{rx}}$ is not revealing, but for which $\pi_{\text{tx}}(h_{\text{tx}})=1$. We define a new transmitter policy $\pi_{\text{tx}}'$ and show that $V^{\pi_{\text{tx}}',\pi_{\text{rx}}}_{s,t}(h_{\text{tx}},h_{\text{rx}})>V^{\pi_{\text{tx}},\pi_{\text{rx}}}_{s,t}(h_{\text{tx}},h_{\text{rx}})$. The new policy is
\begin{equation}
  \pi'_{\text{tx},\tau}(h_{\text{tx}}')=
  \begin{cases}
    \pi_{\text{tx},\tau}(h_{\text{tx}}'), &\tau<t\vee h_{\text{tx}}\not\subset h_{\text{tx}}';\\
    0, &\tau=t;\\
    1, &\tau=t+1;\\
    \pi_{\text{tx},t+1}(h_{\text{tx},0:t+1}'), &\tau=t+2;\\
    \pi_{\text{tx},\tau}(h_{\text{tx}}''), &\text{otherwise,}
  \end{cases}
\end{equation}
where, as in the proof of Theorem~\ref{thm:oneway_cond}, $h_{\text{tx}}''$ is a fictitious history. In this case, it
identical to $h_{\text{tx}}'$ except for step $t$, in which $c''_t=1$, and $c''_{t+1}=\pi_{\text{tx},t+1}(h_{\text{tx},0:t+1}')$. This ensures that the two policies behave identically from step $t+3$ onward.

In other words, policy $\pi'_{\text{tx},\tau}(h_{\text{tx}}')$ forces a transmission at step $t+1$, while its behavior in step $t+2$ mirrors what the original policy would have done in step $t+1$. If $\pi_{\text{tx},t+1}(h_{\text{tx},0:t+1}')=0$, we know that the modified policy always improves the expected value, as it was part of the proof of the general case in Appendix A. On the other hand, if $\pi_{\text{tx},t+1}(h_{\text{tx},0:t+1}')=1$, we need to consider two cases, depending on its action in step $t+2$. As the selected action depends on the outcomes of transmissions at steps $t$ and $t+1$, we cannot compute the probability of these two cases. We thus impose the value improvement condition in both cases.

We first consider the case in which $c_{t+2}=0$: in this case, the original policy transmits at steps $t$ and $t+1$, while the modified policy transmits at steps $t+1$ and $t+2$. The transmission cost is then reduced by $(1-\gamma^2)\beta$ due to the discount. Furthermore, 
since the action of the receiver in step $t+1$ does not affect $s_{t+2}$, the modified policy will provide the receiver with better information about the state in that step, as its transmissions are more recent, and future values will be improved thanks to the Markov property.

In step $t+1$, the modified policy is at a disadvantage if the transmission made by the original policy at time step $t$ succeeds, while its own transmission fails. This happens with probability $\varepsilon(1-\varepsilon)$, and the maximum reward loss is $1$, discounted by $\gamma$. Putting it all together, we get the following condition for $\Delta V$ being positive:
\begin{equation}
    \Delta V\geq(1-\gamma^2)\beta-\gamma\varepsilon(1-\varepsilon).
\end{equation}
Setting $\Delta V>0$ returns the condition in~\eqref{eq:beta_cmab_cond}.

On the other hand, in the case in which $c_{t+2}=1$, the transmission cost is reduced by $\beta$ for the modified policy. However, in addition to the cost at time step $t+1$, we have to consider that it may maintain a disadvantage even after step $t+2$: this occurs if both transmissions made by the modified policy fail, while the extra transmission made at time step $t$ by the original policy is successful. In all other cases, the performance of the two policies from step $t+2$ onward is matched. Following the same strategy as in the previous proof, we set an upper bound to the value difference as $(1-\gamma)^{-1}$. The value difference is then
\begin{equation}
    \Delta V\geq \beta-\gamma\varepsilon(1-\varepsilon)-\gamma^2\varepsilon^2(1-\varepsilon)(1-\gamma)^{-1},
\end{equation}
where the first negative term comes from the possible loss in step $t+1$, while the second comes from losses from step $t+2$ onward. Setting $\Delta V>0$ then yields condition
\begin{equation}
    \beta>\gamma\varepsilon(1-\varepsilon)+\gamma^2\varepsilon^2(1-\varepsilon)(1-\gamma)^{-1}.
\end{equation}
We can easily prove that this is always a looser condition than~\eqref{eq:beta_cmab_cond} through some algebraic manipulation. Repeating this substitution process extends the result to any state $s$ and time step $t$, thus proving the Theorem.

\bibliographystyle{IEEEtran}
\bibliography{biblio}

\end{document}